\documentclass[aps,twocolumn,nofootinbib,floatfix]{revtex4-1}
\usepackage[utf8]{inputenc}
\usepackage{titlesec}
\usepackage[none]{hyphenat}
\usepackage{balance}
\usepackage{amsmath}
\usepackage{amsfonts}
\usepackage{amscd}
\usepackage{amsthm}
\usepackage{xspace}
\usepackage{braket}
\usepackage{graphicx}
\usepackage{listings}
\usepackage{tcolorbox}
\usepackage{pgfplots}
\usepackage{csquotes}
\usepackage{hhline}
\usepackage{amssymb}
\usepackage{titlesec}
\usepackage{braket}
\usepackage{physics}
\usepackage{xcolor}
\usepackage{verbatim}
\usepackage{bm}

\usepackage{dcolumn}
\usepackage{tabularx}
\usepackage{longtable}
\usepackage{braket}
\usepackage[caption=false]{subfig}
\usepackage{float}
\pgfplotsset{compat=1.14}
\usepackage[math]{cellspace}
\usepackage{placeins}
\usepackage[parfill]{parskip}

\newcolumntype{C}{>{\centering\arraybackslash}X}
\titlespacing*{\section}{0pt}{1.1\baselineskip}{\baselineskip}
\usepackage[colorlinks=true,linkcolor=blue,citecolor=blue,urlcolor=blue]{hyperref}

\begin{document}
\title{Quantum Circuit Design Methodology for Multiple Linear Regression}
\thanks{These authors contributed equally to the work:}
\author{Sanchayan Dutta$^{1}$}
\email{sanchayan98@gmail.com}
\author{Adrien Suau$^{2}$}
\email{adrien.suau@cerfacs.fr}
\author{Sagnik Dutta$^{3}$}
\email{sd15ms136@iiserkol.ac.com}
\author{Suvadeep Roy$^{3}$}
\email{sr15ms116@iiserkol.ac.com}
\thanks{\\These authors supervised the project:}
\author{\\Bikash K. Behera$^{3, 4}$}
\email{bikash@bikashsquantum.com}
\author{Prasanta K. Panigrahi$^3$}
\email{pprasanta@iiserkol.ac.in}
\affiliation{$^1$Department of Electronics and Telecommunication Engineering, Jadavpur University, Kolkata 700032, West Bengal, India}
\affiliation{$^2$\normalfont{CERFACS}, \textit{42 Avenue Gaspard Coriolis, 31100 Toulouse, France}}
\affiliation{$^3$\textit{Department of Physical Sciences, Indian Institute of Science Education and Research Kolkata, Mohanpur 741246, West Bengal, India}}
\affiliation{$^4$\textit{ Bikash's Quantum (OPC) Private Limited, Balindi, Mohanpur 741246, Nadia, West Bengal, India}}

\begin{abstract}
\begin{center}\textbf{Abstract}\end{center}
Multiple linear regression assumes an imperative role in \textit{supervised} machine learning. In 2009, Harrow \textit{et al.} [Phys. Rev. Lett. \textbf{103}, 150502 (2009)] showed that their HHL algorithm can be used to \textit{sample} the solution of a linear system $\mathbf{Ax=b}$ exponentially faster than any existing classical algorithm, with some manageable caveats. The entire field of quantum machine learning gained considerable traction after the discovery of this celebrated algorithm. However, effective practical applications and experimental implementations of HHL are still sparse in the literature. Here, we demonstrate a potential practical utility of HHL, in the context of regression analysis, using the remarkable fact that there exists a natural reduction of any multiple linear regression problem to an equivalent linear systems problem. We put forward a $7$-qubit quantum circuit design, motivated from an earlier work by Cao \textit{et al.} [Mol. Phys. \textbf{110}, 1675 (2012)], to solve a $3$-variable regression problem, using only elementary quantum gates. We also implement the Group Leaders Optimization Algorithm (GLOA) [Mol. Phys. \textbf{109} (5), 761 (2011)] and elaborate on the advantages of using such stochastic algorithms in creating low-cost circuit approximations for the Hamiltonian simulation. We believe that this application of GLOA and similar stochastic algorithms in circuit approximation will boost time- and cost-efficient circuit designing for various quantum machine learning protocols. Further, we discuss our Qiskit simulation and explore certain generalizations to the circuit design.
\end{abstract}
\maketitle

\vspace*{-1.2cm}
\section{Introduction}

\label{qhhl_Sec1}
Quantum algorithms running on quantum computers aim at quickly and efficiently solving several important computational problems faster than classical algorithms running on classical computers \cite{qhhl_nielsen, qhhl_Lloyd96, qhhl_Aharonov03, arXivHHL, qhhl_hales2000, qhhl_farhi2014, qhhl_peruzzo2014, qhhl_kandala2017, qhhl_salesman, qhhl_prime, qhhl_grover}. One key way in which quantum algorithms differ from classical algorithms is that they utilize quantum mechanical phenomena such as superposition and entanglement, that allows us to work in exponentially large Hilbert spaces with only polynomial overheads. This in turn, in some cases, allows for exponential speed-ups in terms of algorithmic complexity \cite{qhhl_nielsen}.

In today's world, machine learning is primarily concerned with the development of low-error models in order to make accurate predictions possible by learning and inferring from training data \cite{qhhl_friedman, qhhl_qml}. It borrows heavily from the field of statistics in which linear regression is one of the flagship tools. The theory of multiple linear regression or more generally multivariate linear regression was largely developed in the field of statistics in the pre-computer era. It is one of the most well understood, versatile, and straightforward techniques in any statistician's toolbox. It is also an important and practical \textit{supervised learning} algorithm. Supervised learning is where one has some labeled input data samples $\{\mathbf{x}_i,\mathbf{y}_i\}_{i=1}^{N}$  (where $\mathbf{x}_i$'s are the feature vectors and $\mathbf{y}_i$'s are the corresponding labels) and then based on some criteria (which might depend on the context) chooses a mapping from the input set $\mathbf{X}$ to the output set $\mathbf{Y}$. And that mapping can help to predict the probable output corresponding to an input lying outside of the training data sets. Multiple linear regression is similar in the sense that given some training samples one identifies a closely fitting \textit{hyperplane} depending on the specific choice of a \textit{loss function} (the most common one being a quadratic loss function based on the ``least squares" method). Interestingly, any multiple regression problem can be converted into an equivalent system of linear equations problem or more specifically, a  Quantum Linear Systems Problem (QLSP) problem \cite{qhhl_childs}. The process has been outlined using an example in Section~\ref{qhhl_Sec3}. 

Suppose that we are given a system of $N$ linear equations with $N$ unknowns, which can be expressed as $\mathbf{A}\mathbf{x}=\mathbf{b}$. Now, what we are interested in, is: given a matrix $\mathbf{A} \in \mathbb{C}^{N\times N}$ with a vector $\mathbf{b}\in \mathbb{C}^N$, find the solution $\mathbf{x} \in \mathbb{C}^N$ satisfying $\mathbf{Ax=b}$ (which is $\mathbf{{A}^{-1}b}$ if $\mathbf{A}$ is invertible), or else return a \textit{flag} if no solution exists. This is known as the Linear Systems Problem (LSP). However, we will consider only a special case of this general problem, in form of the Quantum Linear Systems Problem (QLSP) \cite{arXivDervovic,qhhl_childs}.

The quantum version of the LSP problem, the QLSP, can be expressed as:

\noindent
\textit{Let $\mathbf{A}$ be a $N\times N$ Hermitian matrix with a spectral norm bounded by unity and a known condition number $\kappa$. The quantum state on $\lceil{\log N\rceil}$ qubits $\ket{b}$ can be given by} 

\begin{equation}\ket{b} :=\frac{\sum_i b_i\ket{i}}{||\sum_i b_i \ket{i}||}\end{equation}

\textit{and $\ket{x}$ by}

\begin{equation}\ket{x} :=\frac{\sum_i x_i\ket{i}}{||\sum_i x_i \ket{i}||}\end{equation}

\textit{where $b_i,x_i$ are respectively the $\text{i}^{\text{th}}$ component of vectors $\mathbf{b}$ and $\mathbf{x}$. Given the matrix $\mathbf{A}$ (whose elements are accessed by an oracle) and the state $\ket{b}$, an output state $\ket{\tilde{x}}$ is such that 
$||\ket{\tilde{x}}-\ket{x}||_2\leq \epsilon$, with some probability $\Omega(1)$ (practically at least $\frac{1}{2}$) along with a binary flag indicating `success' or `failure'} \cite{qhhl_childs}.

The restrictions on Hermiticity and spectral norm, can be relaxed by noting that, even for a non-Hermitian matrix $\mathbf A$, the corresponding $\smqty[0 & \mathbf{A}^\dagger \\ \mathbf{A} & 0]$ matrix is Hermitian. This implies that we can instead solve the linear system given by $\smqty[0 & \mathbf{A}^\dagger \\ \mathbf{A} & 0]\mathbf{y} = \smqty[\mathbf{b} \\ 0]$, which has the unique solution $\mathbf{y} = \smqty[ 0 \\ \mathbf{x}]$ when $\mathbf{A}$ is invertible \cite{arXivHHL}. Also, any non-singular matrix can be scaled appropriately to adhere to the given conditions on the eigenspectrum. Note that the case when $\mathbf A$ is non-invertible, has already been excluded by the fact that a known finite condition number exists for the matrix $\mathbf{A}$.

In 2009, A. W. Harrow, A. Hassidim, and S. Lloyd \cite{arXivHHL} put forward a quantum algorithm (popularly known as the ``HHL algorithm") to obtain information about the solution $\mathbf{x}$ of certain classes of linear systems $\mathbf{A}\mathbf{x}=\mathbf{b}$.  As we know, algorithms for finding the solutions to linear systems of equations play an important role in engineering, physics, chemistry, computer science, and economics apart from other areas. The HHL algorithm is highly celebrated in the world of quantum algorithms and quantum machine learning, but there is still a distinct lack of literature demonstrating practical uses of the algorithm and their experimental implementations. Experimentally implementing the HHL algorithm for solving an arbitrary system of linear equations to a satisfactory degree of accuracy, remains an infeasible task even today, due to several physical and theoretical restrictions imposed by the algorithm and the currently accessible hardware.  Our fundamental aim in this paper is to put forth a neat technique to demonstrate how HHL can be used for one of the most useful tools in statistical modeling — multiple regression analysis. We show that there is a canonical reduction of any multiple linear regression problem into a quantum linear systems problem (QLSP) that is thereafter amenable to a solution using HHL. It turns out that, after slight modifications and classical pre-processing, the circuit used for HHL can be used for multiple linear regression as well. Here we should point out, in Scott Aaronson's words \cite{qhhl_aaronson}, that the HHL algorithm is mostly a \textit{template} for other quantum algorithms and it has some caveats that must be kept in mind during the circuit design and physical implementation step, to preserve the exponential speedup offered by the algorithm. Nonetheless, these issues are not always detrimental and can often be overcome by appropriately preparing and justifying the circuit design methodology for specific purposes, as we will do here. There is another brief discussion related to these issues in the Conclusions section of the paper.

In Section~\ref{qhhl_Sec4}, we present an application (in the context of multiple regression) of a modified version of the earlier circuit design by Cao \textit{et al.} \cite{qhhl_cao} which was meant for implementing the HHL algorithm for a $4\times 4$ linear system on real quantum computers. This circuit requires only $7$ qubits and it should be simple enough to experimentally verify it if one gets access to a quantum processor having logic gates with sufficiently low error rates. Previously, Pan \textit{et al.} demonstrated HHL on a $4$-qubit NMR quantum computer \cite{qhhl_pan}, so we believe that it will be easily possible to experimentally implement the circuit we discuss, given the rapid rise in the number of qubits in quantum computer chips, in the past few years.

We also note that although HHL solves the QLSP for all matrices $\mathbf{A}$ or $\smqty[0 & \mathbf{A}^\dagger \\ \mathbf{A} & 0]$, it can be efficiently implemented only when they are sparse and well-conditioned (the sparsity condition may be slightly relaxed) \cite{qhhl_childs}. In this context, `efficient' means `at most polylogarithmic in system size'. A $N\times N$ matrix is called \textit{$s$-sparse} if it has at most $s$ non-zero entries in any row or column. We call it simply \textit{sparse} if it has at most $\text{poly}(\log N)$ entries per row \cite{arXivDervovic}. We generally call a matrix well-conditioned when its singular values lie between the reciprocal of its condition number ($\frac{1}{\kappa}$) and $1$ \cite{arXivHHL}. \textit{Condition number} $\kappa$ of a matrix is the ratio of largest to smallest singular value and is undefined when the smallest singular value of $\mathbf{A}$ is $0$. For Hermitian matrices the eigenvalue magnitudes equal the magnitudes of the respective singular values.

At this point, it is important to reiterate that unlike the output $\mathbf{A}^{-1}\mathbf{b}$ of a classical linear system solver, the output copy of $\ket{\tilde{x}}$ does not provide access to the coordinates of $\mathbf{A}^{-1}\mathbf{b}$. Nevertheless, it allows for \textit{sampling} from the solution vectors like $\bra{\tilde{x}} M \ket{\tilde{x}}$, where $M$ is a quantum-mechanical operator. This is one main difference between solutions of the LSP and solutions of the QLSP. We should also keep in mind that reading out the elements of $\ket{\tilde{x}}$ in itself takes $\mathcal{O}(N)$ time. Thus, a solution to QLSP might be useful only in applications where just samples from the vector $\ket{\tilde{x}}$ are needed \cite{arXivHHL,qhhl_childs,arXivDervovic}.

The best existing classical matrix inversion algorithm involves the Gaussian elimination technique which takes $\mathcal{O}(N^3)$ time. For $s$-sparse and positive semi-definite $\mathbf A$, the Conjugate Gradient algorithm \cite{qhhl_hestenes} can be used to find the solution vector $\mathbf{x}$ in $\mathcal{O}(Ns\kappa \log(1/\epsilon))$ time by minimizing the quadratic error function $|\mathbf{Ax-b}|^2$, where $s$ is the matrix sparsity, $\kappa$ is the condition number and $\epsilon$ is the desired precision parameter. On the other hand, the HHL algorithm scales as $\mathcal{O}(\log(N)s^2\kappa^2/\epsilon)$, and is exponentially faster in $N$ but polynomially slower in $s$ and $\kappa$. In 2010, Andris Ambainis further improved the runtime of HHL to $\mathcal{O}(\kappa \log^3 \kappa \log N/\epsilon)$ \cite{qhhl_ambainis}.  The exponentially worse slowdown in $\epsilon$ was also eliminated \cite{qhhl_childs} by Childs \textit{et al.} in 2017 and it got improved to $\mathcal{O} (s\kappa \text{polylog}(s\kappa/\epsilon))$ \cite{arXivDervovic}. Since HHL has logarithmic scaling only for sparse or low-rank matrices, in 2018, Wossnig \textit{et al.} extended the HHL algorithm with quantum singular value estimation and provided a quantum linear system algorithm for dense matrices which achieves a polynomial improvement in time complexity, that is, $\mathcal{O}(\sqrt{N}\text{polylog}(N) \kappa^2/\epsilon)$ \cite{qhhl_wossnig} (HHL retains its logarithmic scaling only for sparse or low-rank matrices). Furthermore, an exponential improvement is achievable with this algorithm if the rank of $\mathbf{A}$ is polylogarithmic in the matrix dimension.

Last but not least, we note that it is assumed that the state $\ket{b}$ can be efficiently constructed i.e., prepared in `polylogarithmic time'. In reality, however, efficient preparation of arbitrary quantum states is hard and is subject to several constraints. 

\renewcommand\thempfootnote{\arabic{mpfootnote}}

\begin{figure*}[t]
    \centering
    \includegraphics[width=\textwidth, trim = 0cm 0.37cm 0cm 0cm, clip]{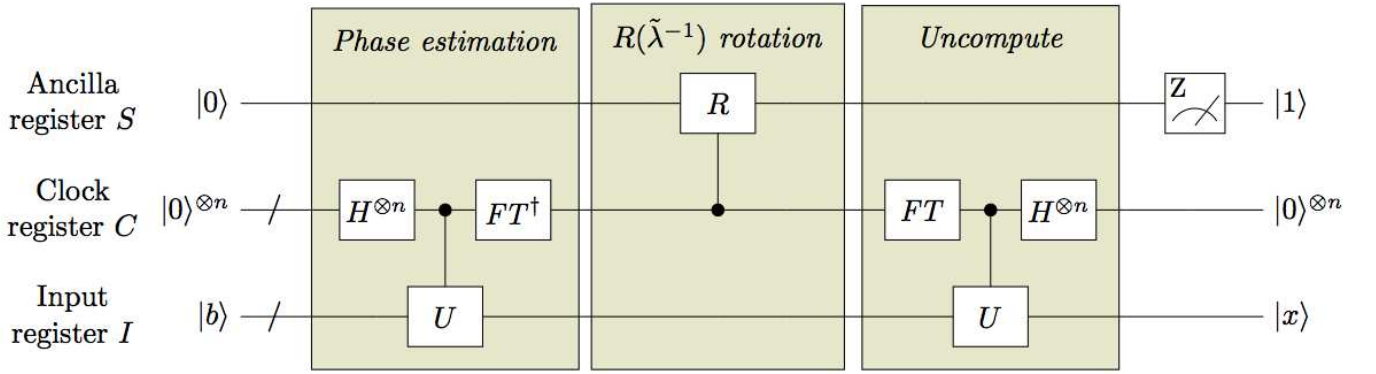}
    \label{qhhl_Fig1}
    \caption{HHL Algorithm Schematic \protect\footnotemark}
\end{figure*}

\footnotetext{The HHL algorithm schematic (Fig.~\ref{qhhl_Fig1}) was generated using the TikZ code provided by Dr. Niel de Beaudrap (Department of Computer Science, Oxford University). It was inspired by figure 5 of the Dervovic \textit{et al.} paper \cite{arXivDervovic}.}

\section{The Harrow Hassidim Lloyd Algorithm}

\label{qhhl_Sec2}
The HHL algorithm consists of three major steps which we will briefly discuss one by one. Initially, we begin with a Hermitian matrix $\mathbf{A}$ and an input state $\ket{b}$ corresponding to our specific system of linear equations. The assumption that $\mathbf{A}$ is Hermitian may be dropped without loss of generality since we can instead solve the linear system of equations given by $\smqty[0 & \mathbf{A}^\dagger \\ \mathbf{A} & 0]\mathbf{y} = \smqty[\mathbf{b} \\ 0]$ which has the unique solution $\mathbf{y}=\smqty[0 \\ \mathbf{x}]$ when $\mathbf{A}$ is invertible. This transformation does not alter the condition number (ratio of the magnitudes of the largest and smallest eigenvalues) of $\mathbf{A}$ \cite{arXivHHL, qhhl_childs}. However, in the case our original matrix $\mathbf{A}$ is not Hermitian, the transformed system with the new matrix $\smqty[0 & \mathbf{A}^\dagger \\ \mathbf{A} & 0]$ needs oracle access to the non-zero entries of the rows \textit{and} columns of $\mathbf{A}$ \cite{qhhl_childs}.
Since $\mathbf{A}$ is assumed to be Hermitian, it follows that $e^{i\mathbf{A}t}$ is unitary. Here $i\mathbf{A}t$ and $-i\mathbf{A}t$ commute and hence $e^{i\mathbf{A}t}e^{-i\mathbf{A}t}=e^{i\mathbf{A}t-i\mathbf{A}t}=e^{\mathbf{0}}=\mathbf{I}$. Moreover, $e^{i\mathbf{A}t}$ shares all its eigenvectors with $\mathbf{A}$, while its eigenvalues are $e^{i\lambda_jt}$ if the eigenvalues of $\mathbf{A}$ are taken to be $\lambda_j$. Suppose that $\ket{u_j}$ are the eigenvectors of $\mathbf{A}$ and $\lambda_j$ are the corresponding eigenvalues. We recall that we assumed all the eigenvalues to be of magnitude less than $1$ (spectral norm is bounded by unity). As the eigenvalues $\lambda_j$ are of the form $0.a_1a_1a_3\cdots$ in binary \cite[p.~222]{qhhl_nielsen}, we will use $\ket{\lambda_j}$ to refer to $\ket{a_1a_2a_3\cdots}$. We know from the spectral theorem that every Hermitian matrix has an orthonormal basis of eigenvectors. So, in this context, $\mathbf{A}$ can be re-written as $\sum_j \lambda_j\ket{u_j}\bra{u_j}$ (via eigendecomposition of $\mathbf{A}$) and $\ket{b}$ as $\sum_j \beta_j \ket{u}_j$.

\subsection{Phase Estimation}
The quantum phase estimation algorithm performs the mapping $\left(\ket{0}^{\otimes n}\right)^C \ket{u}^I \ket{0}^S \mapsto \ket{\tilde \varphi}^C\ket{u}^I\ket{0}^S$ where $\ket{u}$ is an eigenvector of a unitary operator $U$ with an unknown eigenvalue $e^{i2\pi \varphi}$ \cite{qhhl_nielsen}. $\tilde{\varphi}$ is a $t$-bit approximation of $\varphi$, where $t$ is the number of qubits in the clock register. The superscripts on the kets indicate the names of the registers which store the corresponding states. In the HHL algorithm the input register begins with a superposition of eigenvectors instead i.e., $\ket{b} = \sum_j\beta_j\ket{u_j}$ instead of a specific eigenvector $\ket{u}$, and for us the unitary operator is $e^{i\mathbf{A}t}$. So the phase estimation circuit performs the mapping
\begin{equation*}
\left(\ket{0}^{\otimes n}\right)^C \ket{b}^I \mapsto \left(\sum_{j=1}^{N}\beta_j\ket{u_j}^I\ket{\frac{\tilde\lambda_jt_0} {2\pi}}^C\right)
\end{equation*}
where $\tilde\lambda_j$'s are the binary representations of the eigenvalues of $\mathbf{A}$ to a tolerated precision. To be more explicit, here $\tilde{\lambda}_j$ is represented as $b_1b_2b_3\cdots b_t$ ($t$ being number of qubits in the clock register) if the actual binary equivalent of $\lambda_j$ is of the form $\lambda=0.b_1b_2b_3\cdots$. To avoid the factor of $2\pi$ in the denominator, the `evolution time' $t_0$ is generally chosen to be $2\pi$. However, $t_0$ may also be used to `normalize' $\mathbf{A}$ (by re-scaling $t_0$) in case the spectral norm of $\mathbf A$ exceeds $1$\footnote{Ideally, we should know both the upper bound and the lower bound of the eigenvalues, for effective re-scaling. Furthermore, to get accurate estimates, we should attempt to spread the possible values of $\lambda t$ over the whole $2\pi$ range.}. Additionally, an important factor in the performance of the algorithm is the condition number $\kappa$. As $\kappa$ grows, $\mathbf{A}$ tends more and more towards a non-invertible matrix, and the solutions become less and less stable \cite{arXivHHL}. Matrices with large condition numbers are said to be `ill-conditioned'. The HHL algorithm generally assumes that the singular values of $\mathbf{A}$ lie between $1/\kappa$ and $1$, which ensures that the matrices we have to deal with are `well-conditioned'. Nonetheless, there are methods to tackle ill-conditioned matrices and those have been thoroughly discussed in the paper by Lloyd \textit{et al.} \cite{arXivHHL}. It is worth mentioning that in this step the `clock register'-controlled Hamiltonian simulation gate $U$ can be expressed as $\sum_{k=0}^{T-1} \ket{\tau}\bra{\tau}^C \otimes e^{i\mathbf{A}\tau t_0/T}$, where $T=2^t$ and evolution time $t_0 = \mathcal{O}(\kappa/\epsilon)$. Interestingly choosing $t_0 = \mathcal{O}(\kappa/\epsilon)$ can at worse cause an error of magnitude $\epsilon$ in the final state \cite{arXivHHL}.

\subsection[Controlled Rotation]{R$(\tilde{\lambda}^{-1})$ \ \ rotation}

A `clock register' controlled $\sigma_y$-rotation of the `ancilla' qubit produces a normalized state of the form 
\begin{equation*}
\sum_{j=1}^{N}\beta_j\ket{u_j}^I\ket{\tilde\lambda_j}^C\left(\sqrt{1-\frac{C^2}{\tilde\lambda_j^2}}\ket{0}+\frac{C}{\tilde\lambda_j}\ket{1}\right)^S
\end{equation*}
These rotations, conditioned on respective $\tilde{\lambda}_j$, can be achieved by the application of the $\exp(-i\theta\sigma_y) = \left[
\begin{array}{cc}
\cos \theta & - \sin \theta \\
\sin \theta & \cos \theta
\end{array}
\right]$ operators where $\theta = \cos^{-1}\left(\dfrac{C}{\tilde \lambda_j}\right)$. $C$ is a scaling factor to prevent the controlled rotation from being unphysical \cite{arXivDervovic}. That is, practically $C < \lambda_{\text{min}}$ is a safe choice, which may be more formally stated as $C = \mathcal{O}(1/\kappa)$ \cite{arXivHHL}.
    
\subsection{Uncomputation}
In the final step, the inverse quantum phase estimation algorithm sets back the clock register to $(\ket{0}^{\otimes n})^C$ and leaves the remaining state as 
\begin{equation*}
\sum_{j=1}^{N}\beta_j\ket{u_j}^I\left(\sqrt{1-\frac{C^2}{\tilde\lambda_j^2}}\ket{0}+\frac{C}{\tilde\lambda_j}\ket{1}\right)^S
\end{equation*}
Postselecting on the ancilla $\ket{1}^S$ gives the final state $C\sum_{j=1}^{N}(\frac{\beta_j}{\lambda_j})\ket{u_j}^I$  \cite{arXivDervovic}. The inverse of the Hermitian matrix $\mathbf{A}$ can be written as $\sum_j \frac{1}{\lambda_j}\ket{u_j}\bra{u_j}$, and hence $\mathbf{A}^{-1}|b\rangle$ matches $\sum_{j=1}^{N}\frac{\beta_j}{\tilde{\lambda_j}}\ket{u_j}^I$. This outcome state, in the standard basis, is component-wise proportional to the exact solution $\mathbf{x}$ of the system $\mathbf{Ax = b}$ \cite{qhhl_cao}.

\section{Linear Regression Utilizing HHL}

\label{qhhl_Sec3}
Linear regression models a linear relationship between a scalar `response' variable and one or more `feature' variables. Given a $n$-unit data set $\{y_i,x_{i1},\cdots,x_{ip}\}_{i=1}^{n}$, a linear regression model assumes that the relationship between the dependent variable $y$ and a set of $p$ attributes i.e., $\mathbf{x} = \{x_{1},\cdots,x_{p}\}$ is linear \cite{qhhl_friedman}. Essentially, the model takes the form
\begin{equation*}
y_i = \beta_0 + \beta_1 x_1 + \cdots + \beta_p x_{ip} + \epsilon_i = \mathbf{x}_i^T \pmb{\beta} + \epsilon_i    
\end{equation*}
where $\epsilon_i$ is the noise or error term. Here $i$ ranges from $1$ to $n$. $\mathbf{x}_i^T$ denotes the transpose of the column matrix $\mathbf{x}_i$. And $\mathbf{x}_i^T\pmb{\beta}$ is the \textit{inner product} between vectors $\mathbf{x}_i$ and $\pmb{\beta}$. These $n$ equations may be more compactly represented in the matrix notation, as $\mathbf{y=X\pmb{\beta}+\pmb{\epsilon}}$. Now, we will consider a simple example with $3$ feature variables and a bias $\beta_0$. Say our data sets are $\{-\frac{1}{8} + \frac{1}{8\sqrt{2}},-\sqrt{2},\frac{1}{\sqrt{2}},-\frac{1}{2}\}$, $\{\frac{3}{8} - \frac{3}{8\sqrt{2}},-\sqrt{2},-\frac{1}{\sqrt{2}},\frac{1}{2}\}$, $\{-\frac{1}{8} - \frac{1}{8\sqrt{2}},\sqrt{2},-\frac{1}{\sqrt{2}},-\frac{1}{2}\}$ and $\{\frac{3}{8} + \frac{3}{8\sqrt{2}},\sqrt{2},\frac{1}{\sqrt{2}},\frac{1}{2}\}$. Plugging in these data sets we get the linear system:

\begin{equation}\label{qhhl_Eq1}\beta_0 -\sqrt{2} \beta_1 + \frac{1}{\sqrt{2}} \beta_{2} - \frac{1}{2} \beta_3 = -\frac{1}{8} + \frac{1}{8\sqrt{2}}\end{equation}

\begin{equation}\label{qhhl_Eq2}\beta_0 -\sqrt{2} \beta_1 - \frac{1}{\sqrt{2}} \beta_2  + \frac{1}{2} \beta_3 = \frac{3}{8} -\frac{3}{8\sqrt{2}}\end{equation}

\begin{equation}\label{qhhl_Eq3}\beta_0 + \sqrt{2} \beta_1 - \frac{1}{\sqrt{2}} \beta_2  - \frac{1}{2} \beta_3 =  - \frac{1}{8} - \frac{1}{8\sqrt{2}}\end{equation}

\begin{equation}\label{qhhl_Eq4}\beta_0 + \sqrt{2} \beta_1 + \frac{1}{\sqrt{2}} \beta_2  + \frac{1}{2} \beta_3 = \frac{3}{8} + \frac{3}{8\sqrt{2}}\end{equation}

To estimate $\pmb{\beta}$ we will use the popular ``least squares" method, which minimizes the residual sum of squares $\sum_{i=1}^{N}(y_i - \mathbf{x}_i\pmb{\beta}_i)^2$. If $\mathbf{X}$ is \textit{positive definite} (and in turn has \textit{full rank}) we can obtain a unique solution for the best fit $\pmb{\hat{\beta}}$, which is $\mathbf{(X^TX)^{-1}X^Ty}$. It can happen that all the columns of $\mathbf{X}$ are not linearly independent and by extension $\mathbf{X}$ is not full rank. This kind of situation might occur if two or more of the feature variables are perfectly correlated. Then $\mathbf{X^TX}$ would be singular and $\hat{\beta}$ wouldn't be uniquely defined. Nevertheless, there exist techniques like ``filtering'' to resolve the non-unique representations by reducing the redundant features. Rank deficiencies might also occur if the number of features $p$ exceeds the number of data sets $N$. If we estimate such models using ``regularization'', then redundant columns should not be left out. The regulation takes care of the singularities. More importantly, the final prediction might depend on which columns are left out \cite{qhhl_buhlmann}.

Equations \eqref{qhhl_Eq1}-\eqref{qhhl_Eq4} may be expressed in the matrix notation as:
\begin{align}
\left[\begin{matrix}
-\sqrt{2} & 1 & \frac{1}{\sqrt{2}} & -\frac{1}{2} \\
-\sqrt{2} & 1 & -\frac{1}{\sqrt{2}} & \frac{1}{2} \\
-\sqrt{2} & -1 & \frac{1}{\sqrt{2}} & \frac{1}{2} \\
\sqrt{2} & 1 & \frac{1}{\sqrt{2}} & \frac{1}{2}
\end{matrix}\right] \left[\begin{matrix}
\beta_1 \\ \beta_0 \\ \beta_2 \\ \beta_3 \end{matrix}\right] = \left[\begin{matrix}
-\frac{1}{8} + \frac{1}{8\sqrt{2}} \\ \frac{3}{8} - \frac{3}{8\sqrt{2}} \\ \frac{1}{8} + \frac{1}{8\sqrt{2}} \\ \frac{3}{8} + \frac{3}{8\sqrt{2}} \end{matrix}\right]
\end{align}
Note that unlike common convention, our representation of $\mathbf{X}$ does not contain a column full of $1$'s corresponding the bias term. This representation is used simply because of the convenient form that we obtain for $\mathbf{X}^T\mathbf{X}$. The final result remains unaffected as long as $\mathbf{y=X\pmb{\beta}}$ represents the same linear system.

Now 
\begin{align}
\mathbf{X}^T\mathbf{X} = \frac{1}{4} \left[\begin{matrix}
15 & 9 & 5 & -3 \\
9 & 15 & 3 & -5 \\
5 & 3 & 15 & -9 \\
-3 & -5 & -9 & 15
\end{matrix}\right]
\end{align}
and 
\begin{align}
\mathbf{X}^T\mathbf{y} = \begin{bmatrix}
\frac{1}{2} \\ 
\frac{1}{2} \\
\frac{1}{2} \\
\frac{1}{2}
\end{bmatrix}.
\end{align}

Thus we need to solve for $\pmb{\hat{\beta}}$ from $\mathbf{X}^T\mathbf{X}\pmb{\hat{\beta}}=\mathbf{X}^T\mathbf{y}$ \cite{qhhl_friedman}.

\begin{figure*}[t]
\includegraphics[width=\textwidth,trim = 8cm 15cm 2cm 0cm, clip]{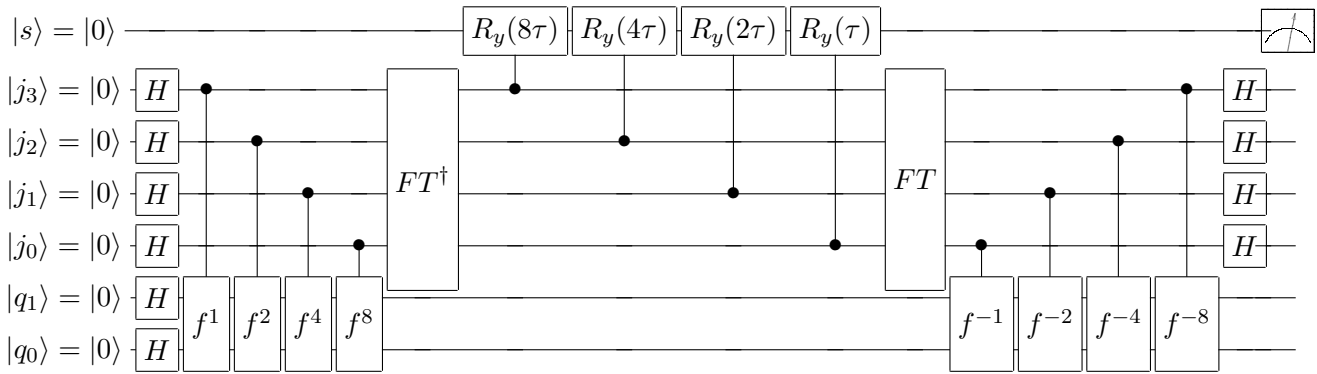}
\caption{Quantum circuit for solving a $4\times 4$ linear system $\mathbf{Ax=b}$. Here $f=\exp(i\mathbf A t/16)$ and $\tau=\pi/2^r$. The top qubit ($|s\rangle$) is the ancilla qubit. The four qubits in the middle ($|j_0\rangle, |j_1\rangle, |j_2\rangle$ and $|j_3\rangle$) stand for the Clock register $C$. The two qubits at the bottom ($|q_0\rangle$ and $|q_1\rangle$) form the Input register $I$ and two Hadamard gates are applied on them to initialize the state $\ket{b}$.}
\label{qhhl_Fig2}
\end{figure*}

\section{Quantum Circuit}
\label{qhhl_Sec4}

Having discussed the general idea behind the HHL algorithm in Section~\ref{qhhl_Sec2} and its possible application in drastically speeding up multiple regression in Section~\ref{qhhl_Sec3}, we now move on to the quantum circuit design meant to solve the $4\times 4$ linear system which we encountered in Section~\ref{qhhl_Sec3} i.e., $\mathbf{X}^T\mathbf{X}\boldsymbol{\hat{\beta}}=\mathbf{X}^T\mathbf{y}$. For sake of convenience we will now denote $\mathbf{X}^T\mathbf{X}$ with $\mathbf{A}$, $\boldsymbol{\hat{\beta}}$ with $\mathbf{x}$ and $\mathbf{X}^T\mathbf{y}$ with $\mathbf{b}$. The circuit requires only $7$ qubits, with $4$ qubits in the ``clock register", $2$ qubits in the ``input register" and the remaining $1$ as an ``ancilla" qubit. At this point it is imperative to mention that we specifically chose the form of the regression data points in the previous Section such that $\mathbf{A}$ turns out to be Hermitian, has four distinct eigenvalues of the form $\lambda_i = 2^{i-1}$ and $\mathbf{b}$ has a convenient form which can be efficiently prepared by simply using two Hadamard gates.

\begin{align}
\mathbf{A} = \mathbf X^T \mathbf X = \frac{1}{4} \left[\begin{matrix}
15 & 9 & 5 & -3 \\
9 & 15 & 3 & -5 \\
5 & 3 & 15 & -9 \\
-3 & -5 & -9 & 15
\end{matrix}\right]
\label{Equation 8}
\end{align}

$\mathbf{A}$ is a Hermitian matrix with eigenvalues $\lambda_1= 1,\lambda_2 = 2, \lambda_3 = 4$ and $\lambda_4 = 8$. The corresponding eigenvectors encoded in quantum states $|u_j\rangle$ may be expressed as 
\begin{equation}
|u_1\rangle = -|00\rangle -|01\rangle - |10\rangle + |11\rangle
\end{equation} 
\vspace*{-5mm}
\begin{equation}
|u_2\rangle = +|00\rangle +|01\rangle - |10\rangle + |11\rangle
\end{equation}
\begin{equation}
|u_3\rangle = +|00\rangle - |01\rangle + |10\rangle + |11\rangle
\end{equation}
\begin{equation}
|u_4\rangle = -|00\rangle + |01\rangle + |10\rangle + |11\rangle
\end{equation}
Also, $\mathbf{b} = \left[\begin{matrix} \frac{1}{2} & \frac{1}{2} & \frac{1}{2} & \frac{1}{2} \end{matrix}\right]^T$ can be written as $\sum_{j=1}^{j=4}\beta_j|u_j\rangle$ where each $\beta_j = \frac{1}{2}$. 

We will now trace through the quantum circuit in Fig.~\ref{qhhl_Fig2}. $|q_0\rangle$ and $|q_1\rangle$ are the input register qubits which are initialized to a combined quantum state 
\begin{equation}
|b\rangle = \frac{1}{2}|00\rangle+\frac{1}{2}|01\rangle+\frac{1}{2}|10\rangle+\frac{1}{2}|11\rangle
\end{equation}
which is basically the state-encoded format of $\mathbf{b}$. This is followed by the quantum phase estimation step which involves a Walsh-Hadamard transform on the clock register qubits $|j_0\rangle,|j_1\rangle,|j_2\rangle,|j_3\rangle$, a clock register controlled unitary gates $U^{2^0},U^{2^1},U^{2^2}$ and $U^{2^3}$ where $U=\exp(i\mathbf{A}t/16)$ and an inverse quantum Fourier transform on the clock register. As discussed in Section~\ref{qhhl_Sec2}, this step would produce the state $\frac{1}{2}|0001\rangle^C|u_1\rangle^I + \frac{1}{2}|0010\rangle^C|u_2\rangle^I+\frac{1}{2}|0100\rangle^C|u_3\rangle^I+\frac{1}{2}|1000\rangle^C|u_4\rangle^I$, which is essentially the same as $\sum_{j=1}^{N}\beta_j\ket{u_j}^I\ket{\frac{\tilde\lambda_jt_0} {2\pi}}^C$, assuming $t_0=2\pi$. Also, in this specific example $|\tilde\lambda_j\rangle = |\lambda_j\rangle$, since the $4$ qubits in the clock register are sufficient to accurately and precisely represent the $4$ eigenvalues in binary. As far as the endianness of the combined quantum states is concerned we must keep in mind that in our circuit $|q_0\rangle$ is the \textit{most significant} qubit (MSQ) and $|q_3\rangle$ is the \textit{least significant} qubit (LSQ).

Next is the $R(\tilde\lambda^{-1})$ rotation step. We make use of an ancilla qubit $|s\rangle$ (initialized in the state $|0\rangle$), which gets phase shifted depending upon the clock register's state. Let's take an example clock register state $|0100\rangle^C = |0\rangle_{q_0}^C\otimes|1\rangle_{q_1}^C\otimes|0\rangle_{q_2}^C\otimes|0\rangle_{q_3}^C$ (binary representation of the eigenvalue corresponding to $|u_3\rangle$, that is $4$). In this combined state, $|q_1\rangle$  is in the state $|1\rangle$ while $|q_0\rangle,|q_2\rangle$ and $|q_3\rangle$ are all the in state $|0\rangle$. This state will only trigger the $R_y(\frac{2\pi}{2^r})$ rotation gate, and none of the other phase shift gates. Thus, we may say that the smallest eigenvalue states in $C$ cause the largest ancilla rotations. Using linearity arguments, it is clear that if the clock register state had instead been $|b\rangle$, as in our original example, the final state generated by this rotation step would be  $\sum_{j=1}^{N}\beta_j\ket{u_j}_I\ket{\tilde\lambda_j}^C((1-C^2/\tilde\lambda_j^2)^{1/2}\ket{0}+\frac{C}{\tilde\lambda_j}\ket{1})^S$ where $C=8\pi/2^r$. For this step, an a priori knowledge of the eigenvalues of $\mathbf{A}$ was necessary to design the gates. For more general cases of eigenvalues, one may refer to \cite{qhhl_cao}.

Then, as elaborated in Section~\ref{qhhl_Sec2}, the inverse phase estimation step essentially reverses the quantum phase estimation step. The state produced by this step, conditioned on obtaining $|1\rangle$ in ancilla is $\frac{8\pi}{2^r}\sum_{j=1}^{j=4}\frac{\frac{1}{2}}{2^{j-1}}|u_j\rangle$. Upon writing in the standard basis and normalizing, it becomes $\frac{1}{\sqrt{340}}(-|00\rangle + 7|01\rangle + 11|10\rangle + 13|11\rangle)$. This is proportional to the exact solution of the system $\mathbf{x} = \frac{1}{32} \left[\begin{matrix}-1 & 7 & 11 & 13\end{matrix}\right]^T$.

\begin{figure*}[t]
\captionsetup[subfigure]{labelformat=empty}
    \centering
    \subfloat[]
    {\includegraphics[width=\textwidth, trim = 3.2cm 18.5cm 0cm 0cm, clip]{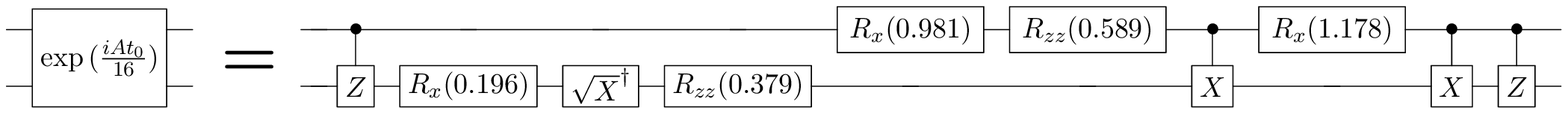}}
    \\[-5ex]
    \subfloat[]
    {\includegraphics[width=\textwidth, trim = 3.3cm 18.5cm 0cm 0cm, clip]{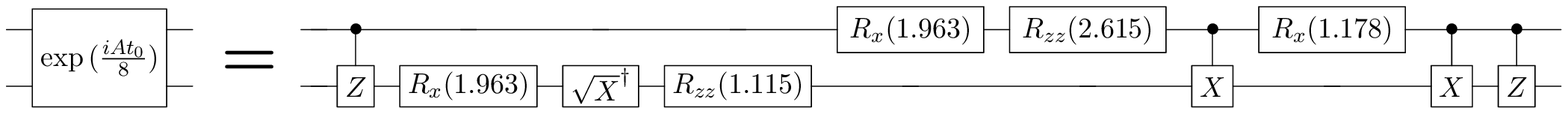}}
    \\[-5ex]
    \subfloat[]
    {\includegraphics[width=\textwidth, trim = 3cm 18.5cm 0cm 0cm, clip]{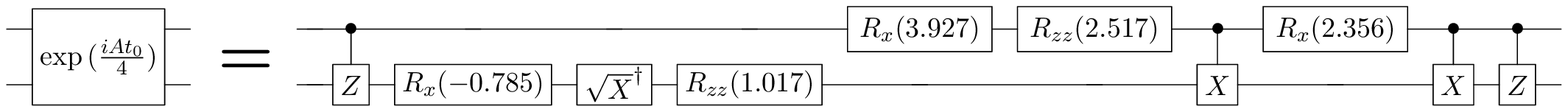}}
    \\[-5ex]
    \subfloat[]
    {\includegraphics[width=\textwidth, trim = 1cm 18.5cm 0cm 0cm, clip]{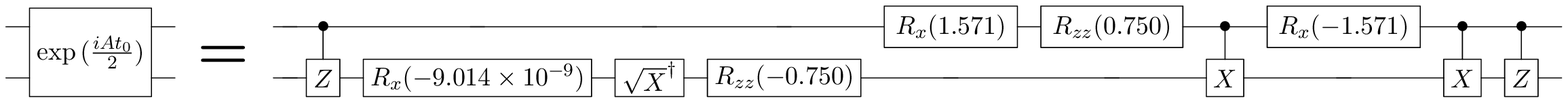}}
    \caption{The Group Leaders Optimization Algorithm is employed to approximately decompose the $\exp(\frac{i\mathbf{A}t}{2^k})$ gates in the Hamiltonian simulation step into elementary quantum gates. The decomposition is not unique. The specific gate decompositions shown in the diagram were taken Cao \textit{et al.}\cite{qhhl_cao} and the angle shifts were corrected using the \textit{scipy.optimize.minimize} module.}
    \label{qhhl_Fig3}
\end{figure*}

\section{Simulation}

We simulated the quantum circuit in Fig.~\ref{qhhl_Fig2} using the Qiskit Aer QasmSimulator backend \cite{qhhl_simulation}, which is a noisy quantum circuit simulator backend. One of the main hurdles while implementing the quantum program was dealing with the Hamiltonian simulation step i.e., implementing the controlled unitary $\mathbf{U} = e^{i\mathbf{A}t}$. Taking a cue from the Cao \textit{et al.} paper \cite{qhhl_cao} which employed the  Group Leaders Optimization Algorithm (GLOA) \cite{qhhl_gloa}, we approximately decomposed the $\mathbf{U}$ gate into elementary quantum gates, as shown in Fig.~\ref{qhhl_Fig3}. It's however important to keep in mind that using GLOA to decompose the $\mathbf{U}$ is useful only when the matrix exponential $e^{i\mathbf{A}t}$ is readily available. Let's call the resulting approximated unitary $\widetilde{\mathbf{U}}$. The parameters of the $R_x$ and $R_{zz}$ gates given in \cite{qhhl_cao} were refined using the \textit{scipy.optimize.minimize} function \cite{qhhl_scipy}, to minimize the Hilbert-Schmidt norm of $\mathbf{U}-\widetilde{\mathbf{U}}$ (which is a measure of the relative error between $\mathbf{U}$ and $\widetilde{\mathbf{U}}$). The \textit{scipy.optimize.minimize} function makes use of the quasi-Newton algorithm of Broyden, Fletcher, Goldfarb, and Shanno (BFGS) \cite{qhhl_BFGS} by default. Also, we noticed that it is necessary to use a controlled-$Z$ gate instead of a single qubit $Z$ gate (as in \cite{qhhl_cao}).

A sample output of the Qiskit code has been shown in the text-box (\ref{qhhl_output}). All the results have been rounded to 4 decimal places. The `Error in found solution' is in essence the 2-norm of the difference between the exact solution and the output solution. We have neglected normalization and constant factors like $\frac{1}{32}$ in the displayed solutions.

\vspace{-4mm}
\begin{center}
\begin{tcolorbox}[width=8.9cm]
\begin{lstlisting}[label={qhhl_output}]
Qiskit Simulation - Sample Output:
(Unnormalized)

    Predicted solution: 
    [-1 7 11 13] 
    
    Simulated experiment solution: 
    [-0.8425 6.9604 10.9980 13.0341]
    
    Error in found solution: 0.1660

\end{lstlisting}
\end{tcolorbox}
\end{center}

\begin{figure}
    \centering
    \includegraphics[width=0.5\textwidth]{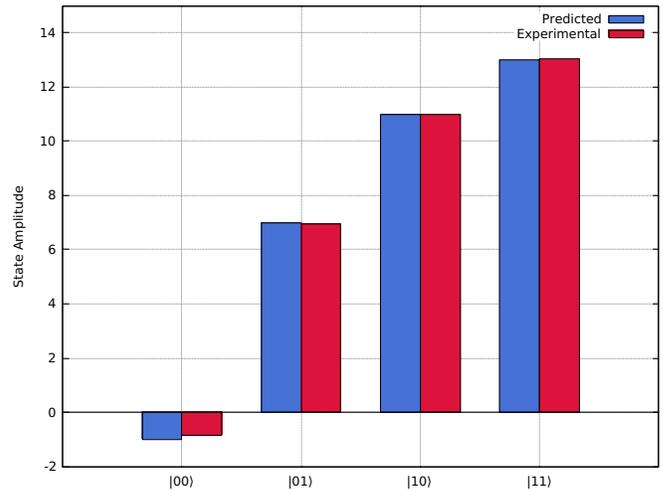}
    \caption{The blue bars represent the predicted state amplitudes for the solution vector $|x\rangle$. The red bars indicate the simulated experiment results using Qiskit Aer QasmSimulator (a noisy quantum simulator backend) for the state amplitudes. The states have not been normalized, so that the deviation from the solution of the original linear equation system is prominent.}
    \label{qhhl_Fig4}
\end{figure}

It is clear from the low error value that the gate decomposition we used for $\widetilde{\mathbf{U}}$ helps to approximately replicate the ideal circuit involving $\mathbf{U}$. Also, the difference between the predicted and simulated results (with shots = $10^5$) has been shown in Fig.~\ref{qhhl_Fig4}.

One should remember that using genetic algorithms like GLOA for the Hamiltonian simulation step is a viable technique only when cost of computing the matrix exponential $e^{i\mathbf{A}t}$ is substantially lower than the cost of solving the corresponding linear system classically. A popular classical algorithm for calculating matrix exponentials is the `The Scaling and Squaring Algorithm' by Al-Mohy and J. Higham \cite{qhhl_mohy, qhhl_higham, qhhl_moler}, whose cost is $\mathcal{O}(n^3)$, and which is generally used along with the Padé approximation by Matlab \cite{qhhl_matlab} and SciPy \cite{qhhl_scipy}. But this algorithm is mostly used for small dense matrices. For large sparse matrices, better approaches exist. For instance, in the Krylov space approach \cite{qhhl_krylov}\cite{qhhl_krylov2}, an Arnoldi algorithm is used whose cost is in the ballpark of $\mathcal{O}(mn)$, where $n$ is the matrix size and $m$ is the number of Krylov vectors which need to be computed. In general, for large sparse matrices, GLOA may be useful but that decision needs to be made on a case-by-case basis depending on the properties of the matrix $\mathbf{A}$.

\section{Discussion and Conclusions}

We have noticed that any multiple linear regression problem can be reduced to an equivalent quantum linear systems problem (QLSP). This allows us to utilize the general quantum speed-up techniques like the HHL algorithm. However, for HHL we need low-error low-cost circuits for the Hamiltonian simulation and controlled rotation steps to get accurate results. There already exist well-defined deterministic approaches like the celebrated Solovay-Kitaev and Trotter-Suzuki algorithms \cite{qhhl_sk, qhhl_ts, qhhl_ts1, qhhl_ts2} that can help to decompose the Hamiltonian simulation unitary $U$ up to arbitrary accuracy. But in most cases, they provide neither minimum cost nor efficiently generable gate sequences for engineering purposes. The Solovay-Kitaev algorithm, which scales as $\mathcal O(\log (1/\epsilon))$, takes advantage of the fact that quantum gates are elements of the unitary group $U(d)$, which is a Lie group and a smooth differentiable manifold, in which one can do precise calculations regarding the geometry and distance between points. Similarly, the Trotter-Suzuki algorithm uses a product formula that can precisely approximate the Hamiltonian exponential, and which exploits the structure of commutation relations in the underlying Lie algebras. In practical engineering scenarios, however, one would prefer a low-cost gate sequence that approximates the unitary just well enough, rather than a high-cost and exact or almost exact decomposition.

This is where stochastic genetic algorithms like GLOA come into play. These are heuristic algorithms that do not have well-defined time complexities and error bounds. Given a particular matrix $\mathbf{A}$, if the time taken to find a circuit approximation for the Hamiltonian evolution, is (say) polynomial-time $\mathcal O(n^c)$ for some constant $c$, the exponential speedup disappears. This remains an issue with GLOA because we still need to classically compute the $2^n \times 2^n$ unitary matrix of the evolution we are trying to approximate. The real advantage of GLOA shows up when dealing with large data sets, as the standard GLOA \cite{qhhl_cao, qhhl_gloa} restricts the number of gates to a maximum of $20$ (a detailed explanation of GLOA is available in the supplementary material). So while the number of linear equations and the size of the data sets can be arbitrarily large, it will only ever use a limited number of gates for approximating the Hamiltonian simulation. This property of GLOA ensures that time taken for Hamiltonian simulation remains nearly independent of the size of the data sets as they grow large. Our circuit design technique will be useful for engineering purposes, particularly when the regression analysis is to be done on extensive amounts of data. Using GLOA, it is also often possible to incorporate other desired aspects of circuit design into the optimization, like specific choices of gate sets. The maximum size of data set on which we can use this approach is only limited by the number of qubits available in the quantum processors.

It is noteworthy that regression analysis is widely used in various scientific and business applications, as it enables one to understand the relationship between variable parameters and make predictions about unknowns. Every day in the actual world, the handling of larger and larger data sets craves for faster and more efficient approaches to such analyses. It is natural to ask whether it might be possible to leverage quantum algorithms for this purpose. The answer seems to be a yes, although there remain several technical and engineering challenges; the dearth of sufficient experimental realizations is testimony to this. Our work on quantum multiple linear regression, which is a first of its kind, shows a practical circuit design method that we believe will pave the path for more viable and economic experimental implementations of quantum machine learning protocols, addressing problems in business analytics, machine learning, and artificial intelligence. In particular, this work opens a new door of possibilities for time- and cost-efficient experimental realizations of various quantum machine learning protocols, exploiting heuristic approaches like GLOA, which will make incrementally large amounts of data tractable.

\section{Acknowledgements}

Sanchayan Dutta would like to thank IISER Kolkata for providing hospitality and support during the period of the Summer Student Research Program (2018). Adrien Suau acknowledges the support provided by CERFACS. Suvadeep Roy and Sagnik Dutta are obliged to INSPIRE, MHRD for their support. B.K.B. acknowledges the financial support of IISER Kolkata. The authors acknowledge the support of the Qiskit SDK in enabling us to simulate the quantum circuits. They appreciate the assistance received from Charles Moussa (Oak Ridge National Lab, Tennessee) in comprehending the intricacies of the GLOA algorithm. They also acknowledge the help received from members of the Quantum Computing Stack Exchange and the Qiskit team in various stages of the research project.

\end{document}


\begin{center}
\textbf{Supplementary Material}
\end{center}

\section{Group Leaders Optimization Algorithm}

The Group Leaders Optimization Algorithm (GLOA) is a genetic algorithm that was developed by Anmer Daskin and Sabre Kais in 2010 \cite{qhhl_gloa}. One of the primary applications of the algorithm is to find low-cost quantum gate sequences to closely approximate any unitary operator \cite{qhhl_daskin_kais_2011}. Generally speaking, the advantage of genetic algorithms compared to other optimization techniques is that they don't get stuck in local extremas. We will briefly discuss the algorithm here, in the context of quantum gate decomposition.

\vspace{0.3cm}
\noindent\textbf{Step I}: For any arbitrary unitary operator $\mathbf{U}_t$, consider a set (of considerable size) of gates out of the basic $R_x, R_y, R_z, R_{zz}$, Pauli-$X$, Pauli-$Y$, Pauli-$Z$, $V$ (square root of Pauli-$X$), $V^{\dagger}$ , $S$ ($\frac{\pi}{8}$ gate), $T$ ($\frac{\pi}{4}$ gate) and $H$ gates along with their controlled counterparts (refer to the Appendix of \cite{qhhl_gloa} for matrix representations of these gates). Which gate set is to be chosen depends on the context and the choice might directly affect the efficiency, precision and convergence time of the algorithm. One usually tries to choose a universal set of quantum gates. Each gate in the chosen gate set is assigned an index from $1$ onwards. For example, if our gate set were $\{V, Z, S, V^{\dagger}\}$ we could have numbered them as $V=1, Z=2, S=3$ and $V^{\dagger}=4$. For this algorithm, any single qubit gate can be represented as a four-parameter string in the form \colorbox{lightgray}{$\langle\text{index number of gate}\rangle\text{ }\langle\text{index of target qubit}\rangle\text{ }\langle\text{index of control qubit}\rangle\text{ }\langle\text{angle of rotation}\rangle$}. The index number of the control and target qubits can vary from $1$ to total number of qubits (on which $\mathbf{U}$ acts) and the angle of rotation can vary from $0$ to $2\pi$. The angle of rotation is represented by positive floating point numbers whereas the indices are represented by natural numbers. The total number of gates in the decomposition is termed as $\text{max}_{\text{gates}}$, which is restricted to a maximum of $20$ in the GLOA. Say the total number of gates in the decomposition is considered to be say $m$, then the corresponding $4m$-parameter string representing the circuit would be of the form \colorbox{lightgray}{$\langle\text{gate}_1\rangle\text{ }\langle\text{t.q}_1\rangle\text{ }\langle\text{c.q}_1\rangle\text{ }\langle\text{angle}_1\rangle$ $\langle\text{gate}_2\rangle\text{ }\langle\text{t.q}_2\rangle\text{ }\langle\text{c.q}_2\rangle\text{ }\langle\text{angle}_2\rangle$ ... $\langle\text{gate}_m\rangle\text{ }\langle\text{t.q}_m\rangle\text{ }\langle\text{c.q}_m\rangle\text{ }\langle\text{angle}_m\rangle$}, where t.q and c.q are abbreviations for `target qubit' and `control qubit' respectively.

\begin{figure}[ht]
    \centering
    \fbox{\includegraphics[width=0.7\textwidth, trim = 0cm 6.5cm 3cm 3cm, clip]{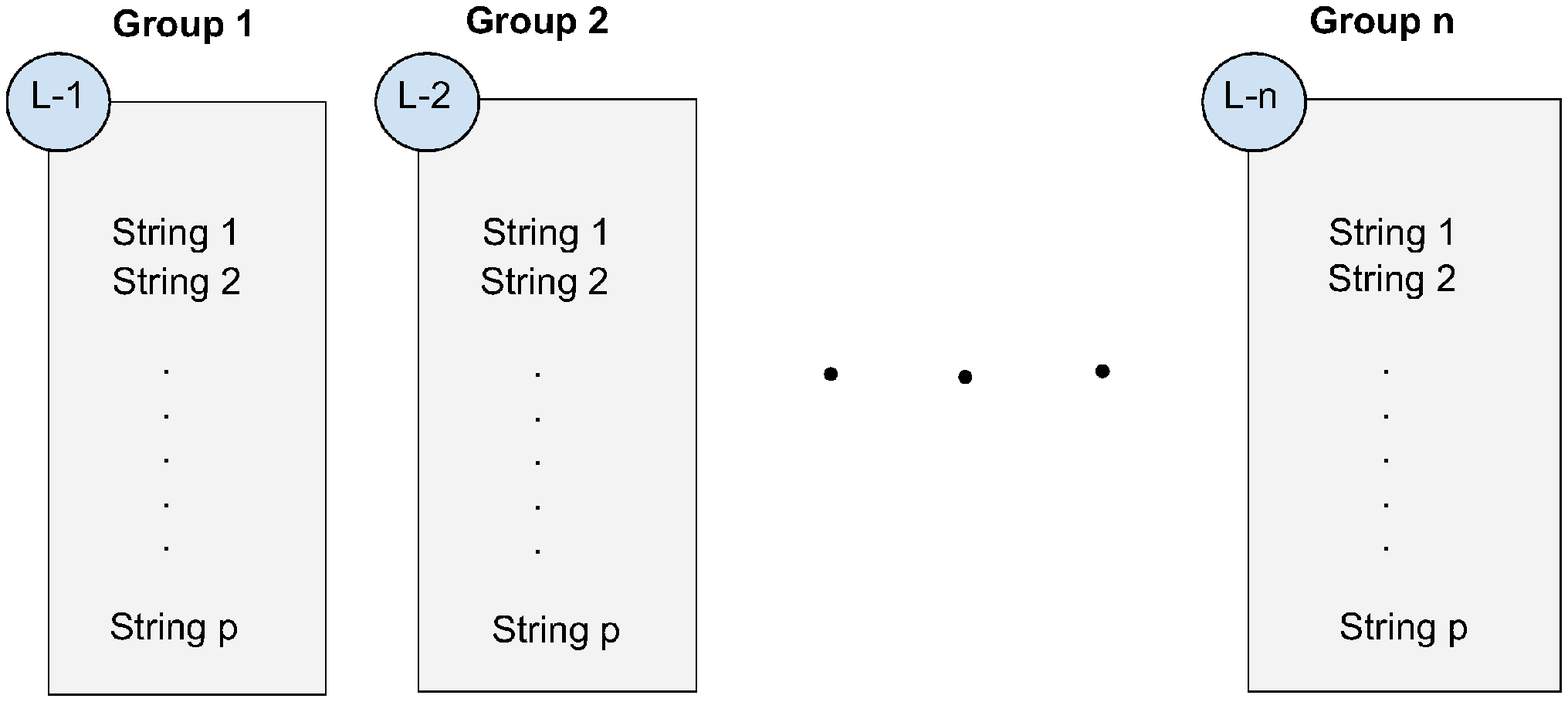}}
    \caption{In Step II of the algorithm $n$ groups of $p$ randomly generated member strings are created. In step III the leader strings L-1, L-2,..., L-n are chosen for each group on basis of their trace fidelities.}
    \label{qhhl_Fig5}
\end{figure}

\vspace{0.3cm}
\noindent\textbf{Step II}: $n$ groups of $p$ such randomly generated $4\times\text{max}_{\text{gates}}$- parameter strings are created. The groups now look as in figure \ref{qhhl_Fig5}. It is important to emphasize that the entire $n\times p$ population of strings is randomly generated, following the constraints on the index numbers (as described in Step I) and angles (the decimal representing an angle can range from $0$ to $2\pi$ only).

\vspace{0.3cm}
\noindent\textbf{Step III}: For the matrix operator equivalent of any such member string i.e., $\mathbf{U}_a$, we can calculate the value of trace fidelity $\mathcal{F}$, which is defined as  
$$\frac{1}{N}|\operatorname{Tr}(\mathbf{U}_t\mathbf{U}_a^{\dagger})|$$ where $\mathbf{U}_t$ is the operator corresponding to the unitary circuit which we are trying to approximate. It is a measure of the \textit{closeness} of the two operators $\mathbf{U}_a$ and $\mathbf{U}_t$. The values of $\mathcal{F}$ can only lie in $[0,1]$. This is because the product of two unitary matrices is always unitary and furthermore all eigenvalues of unitary matrices have a magnitude of $1$ and that the trace of a square matrix is the sum of its eigenvalues. We can clearly see that the higher the value of $\mathcal{F}$, the greater is the closeness of $\mathbf{U}_a$ and $\mathbf{U}_t$. And when $\mathbf{U}_a = \mathbf{U}_t$, the trace fidelity is unity. Using this definition, we calculate the trace fidelities of all the $n\times p$ number of member strings. In each of the $n$ groups, the string having the highest value of trace fidelity is considered to be the leader of that group. So we get $n$ leader strings in each run.  

\vspace{0.3cm}
\noindent\textbf{Step IV}: In this step we mutate every parameter (element) of all the member strings using a weighted sum involving the original string, the leader string and a randomly generated string. Each parameter $pr$ ($1 \leq pr \leq 4\times \text{max}_{\text{gate}}$) of a member string in any group is modified following the rule:
$$\text{new string}_{ij} [pr] = r_1 \times \text{member string}_{ij}[pr] + r_2 \times   \text{leader string}_{i}[pr] + r_3 \times \text{random string}_{ij}[pr].$$ The subscript $i,j$ indicates the $i$-th member of the $j$-th group ($1\leq i \leq n$ and $1\leq j \leq p$). Since the leader string is shared by all members of a particular group, we use only the single index subscript $i$ for it. $\text{random string}_{ij}$ represents a randomly generated string corresponding to the $i$-th member string of the $j$-th group. Here $r_1+r_2+r_3 = 1$ and in general the best results are obtained when $r_1 = 0.8$ and $r_2,r_3$ are considered to be $0.1$ each. We summarize this step with the following pseudo-code:
\\
\begin{tcolorbox}
\begin{lstlisting}[mathescape=true, escapeinside={(*}{*)}, label={qhhl_pseudo_code_1}]
(*\textbf{for}*) $i = 1$ (*\textbf{to}*) $n$ (*\textbf{do}*)
  (*\textbf{for}*) $j = 1$ (*\textbf{to}*) $p$ (*\textbf{do}*)
    (*\textbf{for}*) $pr = 1$ (*\textbf{to}*) $4\times\text{max}_{\text{gate}}$ (*\textbf{do}*)
      $\text{new string}_{ij} [pr] = r_1 \times \text{member string}_{ij}[pr] + r_2 \times   \text{leader string}_{i}[pr] + r_3 \times \text{random string}_{ij}[pr]$
    (*\textbf{end for}*)      
  (*\textbf{if}*) $\text{fidelity}(\text{new string}_{ij})$ $\textbf{is greater than}$ $\text{fidelity}(\text{member string}_{ij})$ $\textbf{then}$ 
    $\text{member string}_{ij} = \text{new string}_{ij}$
  (*\textbf{end if}*)
(*\textbf{end for}*) 
\end{lstlisting}
\end{tcolorbox}
\vspace{0.3cm}
\noindent\textbf{Step V}:
In this step we perform one-way crossovers, also known as parameter transfer. The (randomly chosen) $pr$-th parameter of any random member string $k$ belonging to a group $i$ is replaced with the $pr$-th parameter of $k$ member of a randomly chosen group $x$. It is necessary to keep in mind that only if the trace fidelity of this new string generated after crossover is greater than the original fidelity, the original member string is replaced. This is repeated $t$ times for each group (not for each member). $t$ is a random positive integer bounded above by $\frac{4\times\text{max}_{\text{gate}}}{2}-1$. The pseudo-code is as follows:
\\
\begin{tcolorbox}
\begin{lstlisting}[mathescape=true, escapeinside={(*}{*)}, label={qhhl_pseudo_code_2}]
(*\textbf{for}*) $i = 1$ to $n$ (*\textbf{do}*)
  $t = \text{random}(\frac{4\times\text{max}_{\text{gate}}}{2}-1)$
  (*\textbf{for}*) $j=1$ (*\textbf{to}*) $t$ (*\textbf{do}*)
    $x = \text{random}(n)$ 
    $k = \text{random}(p)$
    $pr = \text{random}(4\times\text{max}_{\text{gate}})$ 
    $\text{new string}_{ik}[pr] = \text{member string}_{xk}[pr]$
    (*\textbf{if}*) $\text{fidelity}(\text{new string}_{ik})$ (*\textbf{is greater than}*) $\text{fidelity}(\text{member string}_{ik})$ (*\textbf{then}*)
      $\text{member string}_{ik} = \text{new string}_{ik}$ 
    (*\textbf{end if}*)
  (*\textbf{end for}*)
(*\textbf{end for}*)
\end{lstlisting}
\end{tcolorbox}
\vspace{0.2cm}
In the pseudo-code, the \textbf{random(x)} function is basically a function which randomly generates a positive integer bounded by $x$. Fast pseudo-random generators such as Xorshift \cite{qhhl_xorshift} or Mersenne twisters \cite{qhhl_mersenne} may be utilized for this purpose. 

\vspace{0.3cm}
\noindent\textbf{Step VI}:
The fidelities are re-calculated for all the strings in all the groups. In case the fidelity of any of the leader strings surpasses the desired fidelity threshold, the algorithm is terminated and that string with the highest fidelity (in the whole population) is returned. We might also terminate the algorithm once a desired number of iterations are completed - say $10,000$. 

\begin{figure}[t]
    \centering
    \fbox{\includegraphics[width=\textwidth, trim = 0cm 4cm 0cm 4cm, clip]{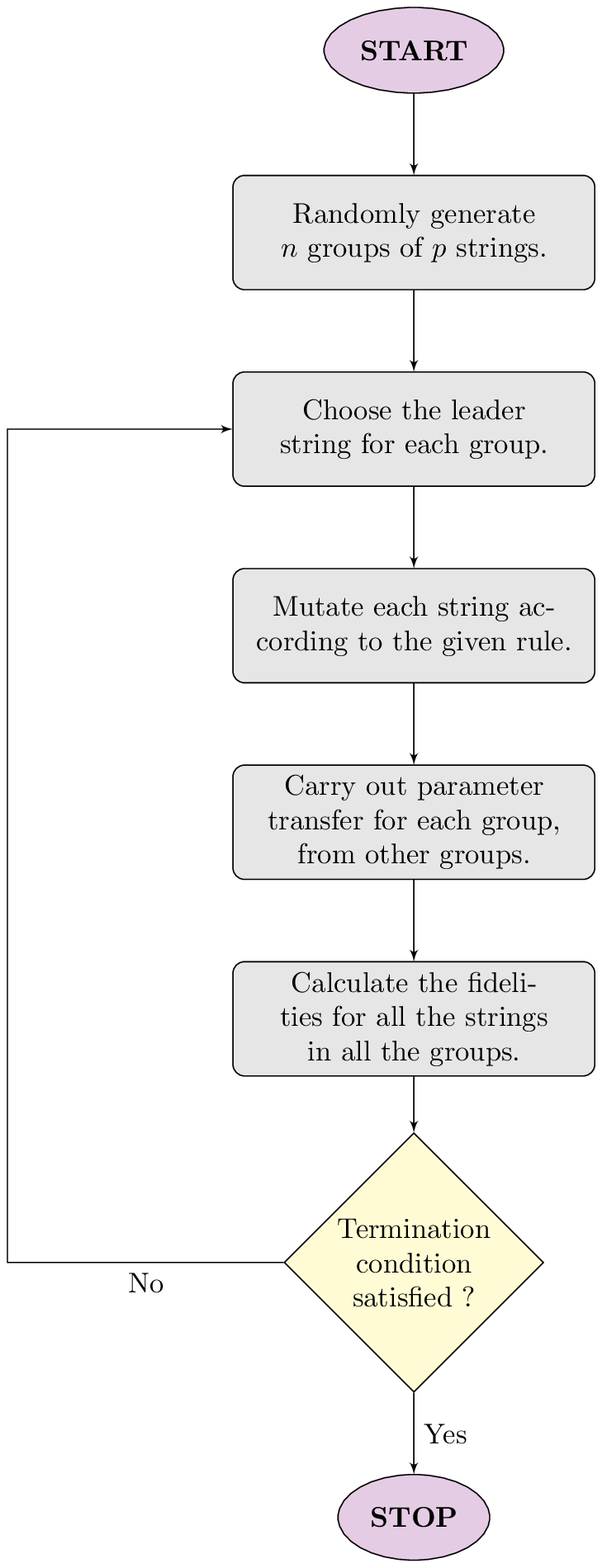}}
    \caption{A flowchart representing the steps of the Group Leaders Optimization algorithm}
    \label{qhhl_Fig6}
\end{figure}
\FloatBarrier

\section{Exact Gate Decomposition of the Hamiltonian Simulation Step}

\begin{figure}[ht]
    \centering
    \subfloat[]
    {\includegraphics[width=0.7\linewidth, trim = 11.5cm 18cm 3cm 1cm,clip]{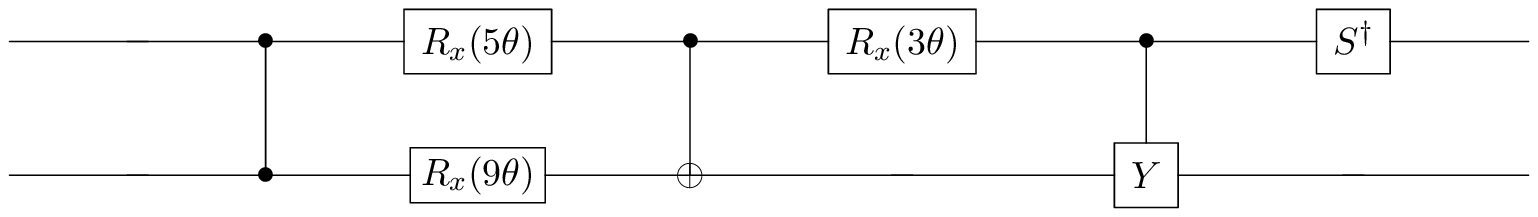}\label{qhhl_Fig7a}
    }
    \\
    \subfloat[]
    {\includegraphics[width=0.7\linewidth, trim = 13cm 18cm 3cm 1cm,clip]{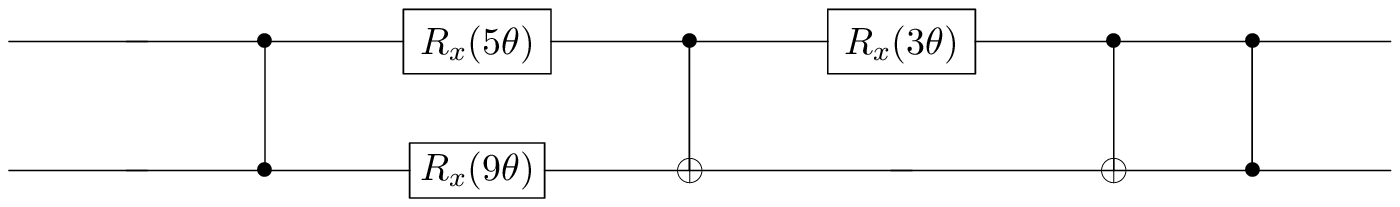}\label{qhhl_Fig7b}
    }
    \caption{(a) shows the exact gate decomposition for $e^{i\mathbf{A}t}$. (b) is a simplified version of (a) obtained by combining the two controlled gates at the end.}
    \label{qhhl_Fig7}
\end{figure}

Any $4\times 4$ two-qubit Hamiltonian, which may be represented as a $4\times 4$ matrix, can be easily decomposed into the four basic Pauli matrices $\sigma_1 = \mathbb{I},\sigma_x = X,\sigma_y = Y,\sigma_z = Z$.
Say we want to express our two-qubit Hamiltonian $\mathbf{H}$ in the form
$$\mathbf{H} = \sum_{i,j=1,x,y,z} a_{i,j}(\sigma_i\otimes \sigma_j).$$
Then the coefficients $a_{i,j}$ turn out to be 
$$a_{i,j}=\frac{1}{4}\operatorname{Tr}[(\sigma_i\otimes\sigma_j)\mathbf{H}].$$ 
The factor of $\frac{1}{4}$ is meant to normalize the Pauli matrices, since $||\sigma_i|| = \sqrt{\operatorname{Tr}[\sigma_i^{\dagger}\sigma_i]} = \sqrt{2}.$
Using this technique, the Pauli decomposition of the $4\times 4$ matrix $\mathbf{A}$ i.e.
\begin{align}
\mathbf{A} = \frac{1}{4} \left[\begin{matrix}
15 & 9 & 5 & -3 \\
9 & 15 & 3 & -5 \\
5 & 3 & 15 & -9 \\
-3 & -5 & -9 & 15
\end{matrix}\right]
\end{align}
turns out to be\footnote{Courtesy of Dr. Alastair Kay (Royal Holloway, University of London), who also designed the quantum circuits in figures \ref{qhhl_Fig7a} and \ref{qhhl_Fig7b}.}

$$\mathbf{A}=\frac{1}{4}(15\mathbb{I}\otimes\mathbb{I}+9Z\otimes X+5X\otimes Z+3Y\otimes Y).$$
Neglecting the scaling factor of $\frac{1}{4}$, we note that each one of the terms commute, which implies
$$e^{i\mathbf{A}\theta}=e^{15i\theta}e^{9i\theta Z\otimes X}e^{5i\theta X\otimes Z}e^{3i\theta Y\otimes Y}.$$Another observation is that the commuting terms are the stabilizers of the 2-qubit cluster state \cite{qhhl_cluster}. So we attempt to use controlled phase gates to get the correct terms. We can rotate the first qubit about the $x$ -axis by an angle $5\theta$ and the second qubit about the $x$-axis by angle $9\theta$. The structure of $e^{3i\theta X\otimes X}$ is a $x$-rotation on the computational basis states $\{|00\rangle,|11\rangle\}$ and another on $\{|01\rangle,|10\rangle\}$. A CNOT gate converts these bases into single qubit bases, controlled off the target qubit. Since both implement the same rotation but controlled off opposite values, we can remove the control. The overall circuit is shown in figure \ref{qhhl_Fig7a}, which can be further simplified by combining the two controlled gates at the end as in figure \ref{qhhl_Fig7b}.